\documentclass[10pt,journal]{IEEEtran}
\usepackage{times}
\usepackage[cmex10]{amsmath}
\usepackage{amssymb}
\usepackage{epsfig,verbatim}
\usepackage{algorithm,algpseudocode} 
\usepackage{hhline}
\usepackage{algpseudocode}
\usepackage{booktabs}
\usepackage{multirow}
\usepackage{graphicx, subfigure}
\usepackage{cite}

\long\def\comment#1{}

\newfont{\bbb}{msbm10 scaled 700}

\newfont{\bb}{msbm10 scaled 1100}
\newcommand{\CC}{\mbox{\bb C}}

\newcommand{\RR}{\mbox{\bb R}}

\newcommand{\av}{{\bf a}}

\newcommand{\ev}{{\bf e}}

\newcommand{\rv}{{\bf r}}
\newcommand{\sv}{{\bf s}}

\newcommand{\xv}{{\bf x}}
\newcommand{\yv}{{\bf y}}
\newcommand{\zv}{{\bf z}}
\newcommand{\zerov}{{\bf 0}}
\newcommand{\onev}{{\bf 1}}

\newcommand{\Hm}{{\bf H}}

\newcommand{\Rm}{{\bf R}}
\newcommand{\Sm}{{\bf S}}

\newcommand{\Cc}{{\cal C}}

\newcommand{\Mc}{{\cal M}}
\newcommand{\Nc}{{\cal N}}

\newcommand{\Rc}{{\cal R}}

\newcommand{\alphav}{\hbox{\boldmath$\alpha$}}
\newcommand{\betav}{\hbox{\boldmath$\beta$}}

\newcommand{\Lambdam}{\hbox{\boldmath$\Lambda$}}

\renewcommand{\arg}{{\hbox{arg}}}

\renewcommand{\Re}{{\rm Re}}
\renewcommand{\Im}{{\rm Im}}
\newcommand{\eqdef}{\stackrel{\Delta}{=}}

\newcommand{\transp}{{\sf T}}

\usepackage{xcolor}

\newtheorem{definition}{Definition}

\title{Construction of One-Bit Transmit-Signal Vectors for Downlink MU-MISO Systems with PSK Signaling}

\author{
\IEEEauthorblockN{
              Gyu-Jeong Park and Song-Nam Hong}\\
\IEEEauthorblockA{Ajou University, Suwon, Korea,\\
              email: \{net2616, snhong\}@ajou.ac.kr}
}

\begin{document}

\maketitle

\date{}


\begin{abstract}
We study a downlink multi-user multiple-input single-output (MU-MISO) system in which the base station (BS) has a large number of antennas with cost-effective one-bit digital-to-analog converters (DACs). In this system, we first identify that antenna-selection can yield a non-trivial symbol-error-rate (SER) performance gain by alleviating an error-floor problem. Likewise the previous works on one-bit precoding, finding an optimal transmit-signal vector (encompassing precoding and antenna-selection) requires exhaustive-search due to its combinatorial nature. Motivated by this, we propose a low-complexity two-stage algorithm to directly obtain such transmit-signal vector. In the first stage, we obtain a {\em feasible} transmit-signal vector via iterative-hard-thresholding algorithm where the resulting vector ensures that each user's noiseless observation is belong to a desired decision region. In the second stage, a bit-flipping algorithm is employed to refine the feasible vector so that each user's received signal is more robust to additive Gaussian noises. Via simulation results, we demonstrate that the proposed method can yield a more elegant performance-complexity tradeoff than the existing one-bit precoding methods.
\end{abstract}

\section{Introduction}

Massive multiple-input multiple-output (MIMO) is one of the promising techniques to cope with the predicted wireless data traffic explosion \cite{Marzetta2010,Adihikary2013,Masouros2013,Lu2014}. In downlink massive MIMO systems, it was shown that low-complexity linear precoding methods as zero-forcing (ZF) and regularized ZF (RZF) achieve an almost optimal performance \cite{Peel2005}. In contrast, the use of a large number of antennas considerably increases the hardware cost and the radio-frequency (RF) circuit consumption \cite{Yang2013}. Hybrid analog-digital precoding (a.k.a., hybrid precoding) is one of the promising methods to address the above problems since it can reduce the number of RF chains \cite{Molisch2017, Alkhateeb2014, Mo2017}. An alternative method is to make use of low-resolution digital-to-analog converters (DACs) (e.g., 1$\sim$3 bits), which can greatly reduce the cost and power consumption per RF chain \cite{Yang2013}. Specifically, in this case, each antenna's transmit symbol is equivalent to a constant-envelop symbol, which enables the use of low-cost power amplifiers and thus reduces the hardware complexity.


Because of the potential merits of using one-bit DACs, there have been numerous studies on the precoding methods for such downlink massive MIMO systems 
\cite{Jacobsson2017, Saxena2017, Usman2016, Landau2017, Castaneda2017, Li2018}. In \cite{Jacobsson2017}, various non-linear precoding techniques were proposed based on semidefinite relaxation (SDR), squared $\ell_{\infty}$-norm relaxation, and sphere decoding. The authors proposed low-complexity {\em quantized} precoding methods as quantized ZF (QZF) \cite{Saxena2017} and quantized minimum-mean squared error (QMMSE) \cite{Usman2016}, which simply applied the one-bit quantization to the outputs of the conventional linear precoding methods. Also, a branch-and-bound and a biconvex relaxation approaches were presented in  \cite{Landau2017} and \cite{Castaneda2017}, respectively. However, these methods either suffer from a severe performance loss or require an expensive computational complexity (see \cite{Li2018} for more details). Very recently, focusing on phase-shift-keying (PSK) constellations,  a low-complexity symbol-scaling method was proposed in \cite{Li2018}, where it optimizes a transmit-signal vector directly in an efficient sequential fashion as a function of an instantaneous channel matrix and users' messages (called symbol-level operation). Also, it was shown that the symbol-scaling method can yield a comparable performance to the previous non-linear precoding methods with a much lower computational complexity.

\begin{figure}
\centerline{\includegraphics[width=9cm]{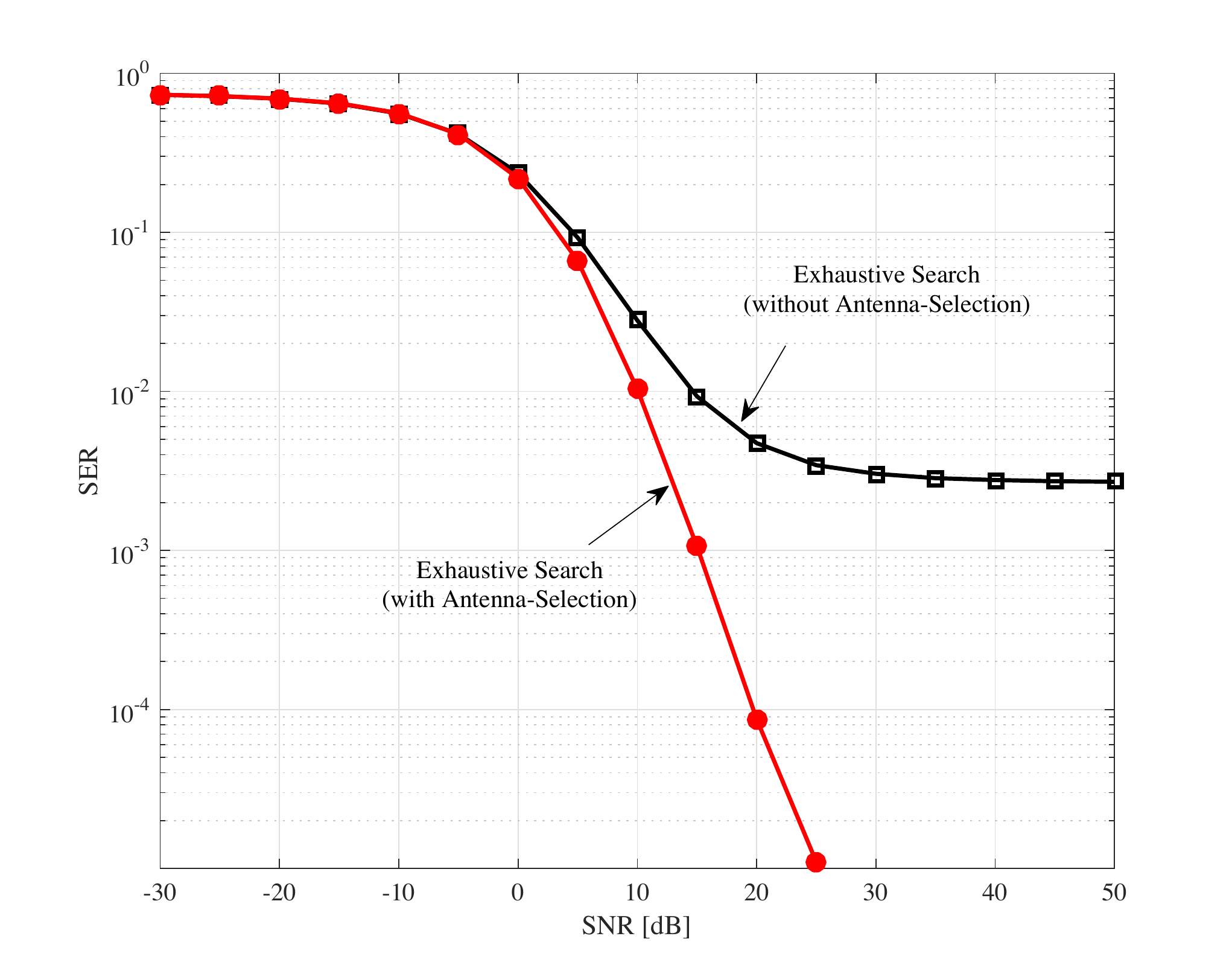}}
\caption{$N_{\rm t}=8$ and $K=2$. Performance improvement of antenna-selection in one-bit precoding for downlink MU-MIMO systems with one-bit DACs.}
\label{FullSearch}
\end{figure}

{\bf Our contributions:} In this paper, we study a downlink MU-MISO system with one-bit DACs. We first identify that antenna-selection can yield a non-trivial bit-error-rate (BER) performance gain by alleviating error-floor problems (see Fig.~\ref{FullSearch}). Clearly, this performance gain is because antenna-selection can enlarge the set of possible transmit-signal vectors (i.e., search-space) compared with the previous precoding methods in \cite{Jacobsson2017, Saxena2017, Usman2016, Landau2017, Castaneda2017, Li2018}. However, as in conventional one-bit precoding optimization, finding an optimal transmit-signal vector (encompassing precoding and antenna-selection) requires exhaustive-search due to its combinatorial nature. Motivated by this, we propose a low-complexity algorithm to solve the above problem (i.e., joint optimization of precoding and antenna-selection), which consists of the following two stages. In the first stage, we obtain a {\em feasible} transmit-signal vector via iterative-hard-thresholding (IHT) algorithm where the resulting vector guarantees that each user's noiseless observation is belong to a desired decision region. Namely, it can improve the BER performances at high-SNR regimes (i.e., error-floor regions) by lowering an error-floor.  In the second stage, we refine the above transit-signal vector using a bit-flipping (BF) algorithm so that each user's received signal is more robust to additive Gaussian noises. In other words, it can improve the BER performances at low-SNR regimes (i.e., waterfall regions). Finally, we provide simulation results to demonstrate that the proposed method can improve the performance of the existing symbol-scaling method in \cite{Li2018} with a comparable computational complexity.


The rest of paper is organized as follows. In Section~\ref{system}, we provide some useful notations and describe the system model. In Section~\ref{sec:main}, we propose a low-complexity algorithm to optimize a transmit-signal vector directly for downlink MU-MISO systems with one-bit DACs. Simulation results are provided in~\ref{sec:SIM}. Section~\ref{sec:con} concludes the paper.

\section{Preliminaries}\label{system}

In this section, we provide some useful notations and describe the system model.

\subsection{Notations}\label{subsec:notation}
The lowercase and uppercase boldface letters represent column vectors and matrices, respectively, and $(\cdot)^{\transp}$ denotes the conjugate transpose of a vector or matrix.
For any vector $\xv$, $x_i$ represents the $i$-th entry of $\xv$. Let  $[a:b]\eqdef\{a,a+1,...,b\}$ for non-negative integers $a$ and $b$ with $a < b$. Similarly, let $[b]=\{1,...,b\}$ for any positive integer $b$.  $\Re(\av)$ and $\Im(\av)$ represent its real and complex part of  a complex vector $\av$, respectively.  Also, we define a natural mapping $g(\cdot)$ which maps a complex value into a real-valued vector, i.e., for each  $x \in \CC$, we have that
\begin{equation}\label{eq:real-complex}
g(x) = \left[\Re(x), \Im(x)\right]^{\transp}.
\end{equation} The inverse mapping of $g$ is denoted as $g^{-1}$. If $g$ or $g^{-1}$ is applied to a vector or a set, we assume they operate element-wise. For example, we have that
\begin{equation}
g(\left[x_1, x_2\right]^{\transp})=\left[\Re(x_1), \Im(x_1), \Re(x_2), \Im(x_2)\right]^{\transp}.
\end{equation} Also, for any complex-value $x \in \CC$, we define a real-valued matrix expansion $\phi(x)$ as
\begin{equation}
\phi(x) =  \begin{bmatrix}
\Re(x) & - \Im(x)\\
\Im(x) &  \Re(x)
\end{bmatrix}.
\end{equation} Finally, we let $\Rm(\theta)$ denote a rotation matrix with a parameter $\theta$ as 
\begin{equation*}
\Rm(\theta) = \begin{bmatrix}
\cos\theta &  -\sin\theta\\
\sin\theta & \cos\theta
\end{bmatrix},
\end{equation*} which rotates the following column vector in the counterclockwise through an angle $\theta$ about the origin.

\subsection{System Model}\label{subsec:model}

\begin{figure}
\centerline{\includegraphics[width=7cm]{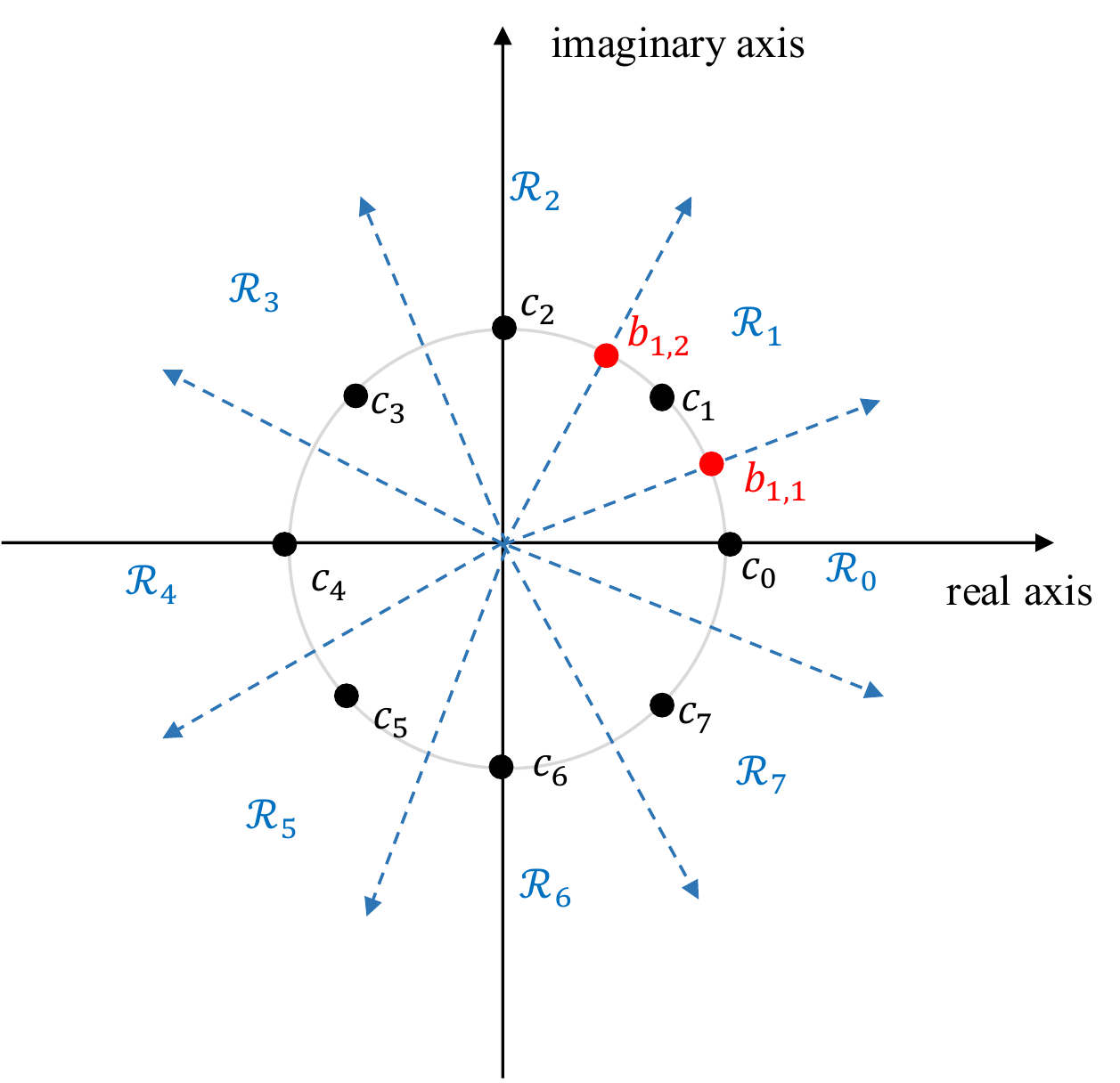}}
\caption{8-PSK constellation in complex domain and the corresponding decision regions. $\Rc_{i}$ represents the decision region of $c_i \in \Mc$.}
\label{8PSK}
\end{figure}

We consider a single-cell downlink MU-MISO system in which one BS with $N_{\rm t}$ antennas communicates with single-antenna $K \ll N_{\rm t}$ users simultaneously in the same time-frequency resources. Focusing on the impact of one-bit DACs in the transmit-side operations, it is assumed that the BS is equipped with one-bit DACs while each user (receiver) is with ideal analog-to-digital converters (ADCs) with infinite resolution. As in the closely related works \cite{Saxena2017,Usman2016,Li2018}, we also consider a normalized $m$-PSK constellation $\Cc=\{c_0,c_1,...,c_{m-1}\}$, each of which constellation point is defined as
\begin{equation}
c_i = \cos\left(2\pi i/m\right) + j \sin\left(2\pi i/m \right) \in \CC.
\end{equation} The constellation points of $8$-PSK are depicted in Fig.~\ref{8PSK}. Let $\mu_{k} \in [0:m-1]$ be the user $k$'s message for $k \in [K]$ and also  let $\xv=\begin{bmatrix}x_1,&...,&x_{N_{\rm t}}\end{bmatrix}^{\transp}$ be a transmit-signal vector at the BS. Under the use of one-bit DACs and antenna-selection, each $x_i$ can be chosen with the restriction of 
\begin{equation}
\Re(x_i) \mbox{ and } \Im(x_i) \in \{-1,0,1\}.
\end{equation} This shows that, compared with the related works \cite{Saxena2017,Usman2016,Li2018}, antenna-selection can enlarge the set of possible symbols (i.e., search-space) per real (or imaginary) part of each antenna. Then, the received signal vector at the $K$ users is given by
\begin{equation}\label{input-output}
\yv= \sqrt{\rho}\Hm\xv+ \zv,
\end{equation} where $\Hm \in \CC^{K\times N_{\rm t}}$ denotes the flat-fading Rayleigh channel with each entry following a complex Gaussian distribution and $\zv \in \CC^{K\times 1}$ denotes the additive Gaussian noise vector whose elements are distributed as circularly symmetric complex Gaussian random variables with zero-mean and unit-variance, i.e., $z_i \sim \Cc\Nc(0,1)$. Also, $\rho$ is chosen according to the per-antenna power constraint.  For simplicity, we assume the uniform power allocation for the antenna array. 

In this system, our purpose is to develop a low-complexity algorithm to optimize a transmit-signal vector $\xv$ (i.e., joint optimization of precoding and antenna-selection) with the assumption that the BS is aware of a perfect channel state information (CSI), which will be provided in Section~\ref{sec:main}. Before explaining our main result, we provide the following definition which will be used throughout the paper.
\begin{definition}\label{def:DR} {\em (Decision Regions)} For each $c_i \in \Cc$, a {\em decision region} $\Rc_{i}$ is defined as
\begin{equation}
\Rc_{i} \eqdef \left\{y \in \CC: |y - c_i| \leq \min_{j\in[0:m-1]:j\neq i}|y - c_j| \right\}.
\end{equation} If the user $k$ receives a $y_k \in \Rc_{\mu_k}$, then it decides the decoded message as $\mu_k \in [0:m-1]$. \hfill$\blacksquare$
\end{definition}

\section{The Proposed Transmit-Signal Vectors}\label{sec:main}

In this section, we derive a mathematical formulation to optimize a transmit-signal vector (encompassing precoding and antenna selection) for downlink MU-MISO systems with one-bit DACs, and then present a low-complexity algorithm to solve such problem efficiently.

For the ease of explanation, we first introduce the equivalent real-valued representation of the complex input-output relationship in (\ref{input-output}), which is given by
\begin{equation}
\tilde{\yv}=\sqrt{\rho}\tilde{\Hm}\tilde{\xv} + \tilde{\zv},
\end{equation}  where $\tilde{\yv} = g(\yv)$, $\tilde{\xv}=g(\xv)$, $\tilde{\zv}=g(\zv)$, and $\tilde{\Hm}$ denotes the $2K\times 2N_{\rm t}$ real-valued matrix which is obtained by replacing $h_{i,j}$ (e.g., the $(i,j)$-th entry of $\Hm$) with the $2\times 2$ matrix $\phi(h_{i,j})$ for all $i,j$. For the resulting model, we can define the real-valued constellation $\tilde{\Cc}=\{g(c_0), g(c_1),...,g(c_{m-1})\}$ where
\begin{equation}
g(c_i)=[\cos(2\pi i/m), \sin(2\pi i/m)]^{\transp}.
\end{equation} From Definition~\ref{def:DR}, the decision region in $\RR^2$ for each $g(c_i)$ is simply obtained  as
\begin{equation}
\tilde{\Rc}_{i} = g(\Rc_{i}).
\end{equation} Furthermore, as shown in Fig.~\ref{8PSK}, each $\tilde{\Rc}_{i}$ can be represented as linear combination of two basis vectors $\sv_{i,1}$ and $\sv_{i,2}$ as 
\begin{equation}
\tilde{\Rc}_{i} \eqdef \left\{\alpha_{i,1} \sv_{i,1} + \alpha_{i,2} \sv_{i,2}: \alpha_{i,1},\alpha_{i,2} > 0\right\},
\end{equation} where the basis vectors are easily obtained using rotation matrices such as
\begin{equation*}
\sv_{i,\ell}=
\Rm\left(\frac{\pi(-1)^{\ell}}{m}\right)\begin{bmatrix}
 \cos\left(\frac{2\pi i}{m}\right)\\
 \sin\left(\frac{2\pi i}{m}\right)
\end{bmatrix}=\begin{bmatrix}
 \cos\left(\frac{\pi (2i+(-1)^{\ell})}{m}\right)\\
 \sin\left(\frac{\pi (2i+(-1)^{\ell})}{m}\right)
\end{bmatrix},
\end{equation*} for $\ell=1,2$. Using two basis vectors, we define the $2\times 2$ real-valued matrix $\Sm_i \eqdef \begin{bmatrix}\sv_{i,1} & \sv_{i,2} \end{bmatrix}$.
Since $\Sm_i$ has full-rank, the inverse matrix of $\Sm$ exists and is easily computed as
\begin{equation}\label{inverse:Si}
\Sm_{i}^{-1} = \frac{1}{\sin(2\pi/m)}
\begin{bmatrix}
 \sin\left(\frac{\pi (2i+1)}{m}\right) & -\cos\left(\frac{\pi (2i+1)}{m}\right)\\
 -\sin\left(\frac{\pi (2i-1)}{m}\right) &  \cos\left(\frac{\pi (2i-1)}{m}\right)
\end{bmatrix}.
\end{equation}

We are now ready to explain how to optimize a transmit-signal vector $\xv$ efficiently. For simplicity, we let $\rv = \Hm\xv \in \CC^{2K\times 1}$ denote the {\em noiseless} received vector at the $K$ users. Regarding our optimization problem, we first provide the following key observations:

\begin{figure}
\centerline{\includegraphics[width=5.5cm]{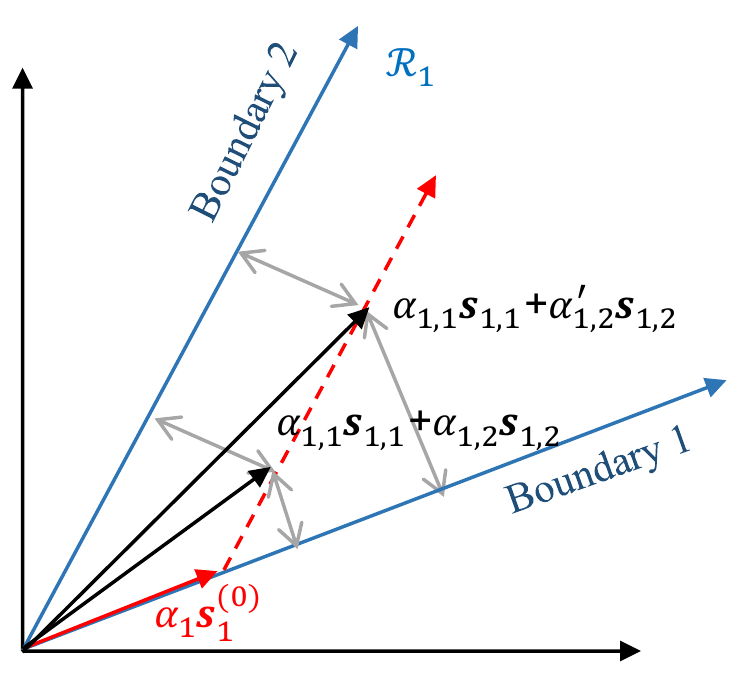}}
\caption{The decision region $\Rc_{1}$  in $\RR^2$ and the impact of larger coefficients.}
\label{Bound}
\end{figure}

\begin{itemize}
\item {\bf (Feasibility condition)} To ensure that all the $K$ users recover their own messages, a transmit-signal vector $\xv$ should be constructed such that 
\begin{equation}\label{criteria0}
r_k \in \Rc_{\mu_k} (\mbox{equivalently}, g(r_k) \in \tilde{\Rc}_{\mu_k}),
\end{equation} for $k\in [K]$. Accordingly, $g(r_k)$ should be represented as 
\begin{equation}\label{criteria1}
g(r_k) = \alpha_{k,1} \sv_{\mu_k,1} + \alpha_{k,2} \sv_{\mu_k,2},
\end{equation} for some positive coefficients $\alpha_{k,1}$ and  $\alpha_{k,2}$. This is called feasibility condition and a vector $\xv$ to satisfy this condition called {\em feasible} transmit-signal vector.
\item {\bf (Noise robusteness)} The condition in (\ref{criteria0}) cannot guarantee good performances in practical SNR regimes (e.g., waterfall regions) due to the impact of additive Gaussian noises. Thus, we need to refine  the above feasible transmit-signal vector so that $\alpha_{k,1}$ and  $\alpha_{k,2}$ are maximized.
\end{itemize} 
We will formulate an optimization problem mathematically which can find a transmit-signal vector $\xv$ to satisfy the above requirements.  From (\ref{criteria1}), we can express the feasibility condition in a matrix form:
\begin{equation}\label{eq:equality}
g(\rv)=\tilde{\Hm}\tilde{\xv}=\Sm\alphav,
\end{equation} where $\alphav = [ \alpha_{1,1}, \alpha_{1,2}, \cdots, \alpha_{K,1}, \alpha_{K,2}]^{\transp}$
and $\Sm=\mbox{diag}(\Sm_{\mu_1},\cdots,\Sm_{\mu_{K}})$ denotes the block diagonal matrix having the $i$-th diagonal block $\Sm_{\mu_{i}}$.
From the block diagonal structure and (\ref{inverse:Si}), we can easily obtain the inverse matrix of $\Sm$ as 
\begin{equation}
\Sm^{-1}=\mbox{diag}(\Sm_{\mu_1}^{-1},\cdots,\Sm_{\mu_{K}}^{-1}).
\end{equation} Then,  the feasibility condition in (\ref{eq:equality}) can be rewritten as
\begin{equation}
\alphav = \Lambdam \tilde{\xv}.
\end{equation} where $\Lambdam \eqdef \Sm^{-1}\tilde{\Hm} \in \RR^{2K\times 2N_{\rm t}}$. Note that  $\Lambdam$ is a known matrix since it is completely determined from the channel matrix $\Hm$ and users' messages $(\mu_1,...,\mu_K)$. Taking the feasibility condition and noise robustness into account, our optimization problem can formulated as
\begin{align}
\max_{\tilde{\xv}} &\min\{\alpha_{k,i}: k\in [K], i=1,2\}\label{obj}\\
\mbox{subject to }&\alphav = \Lambdam \tilde{\xv}\label{const1}\\
& \alpha_{k,1}, \alpha_{k,2} > 0, k \in [K] \label{const2}\\
&\tilde{\xv}\in \{-1,0,1\}^{2N_{\rm t}}.\label{const3}
\end{align} 
For the above optimization problem, the objective function aims to maximize the minimum of positive coefficients $\alpha_{k,i}$'s. This is motivated by the fact that a larger value of the coefficients $\alpha_{k,i}$'s yields a larger distance to the other decision regions. As an example, consider the decision region $\Rc_1$ as illustrated in Fig.~\ref{Bound}.  Obviously, $\alpha_{1,1} \sv_{1,1} + \alpha_{1,2}' \sv_{1,2}$ has a larger distance from the boundary 1 than $\alpha_{1,1} \sv_{1,1} + \alpha_{1,2} \sv_{1,2}$ while both have the same distance from the boundary 2. Likewise, if $\alpha_{1,1}$ increases for a fixed $\alpha_{1,2}$, then the distance from the boundary 2 increases by keeping the distance from the boundary 1.  In order to increase the distance from the other decision regions, therefore, it would make sense to maximize the minimum of two coefficients $\alpha_{1,1}$ and $\alpha_{1,2}$. By extending this to all the $K$ users, we can obtain the objective function in  (\ref{obj}). 

Unfortunately, finding an optimal solution to the above optimization problem is too complicated due to its combinatorial nature. We thus propose a low-complexity two-stage algorithm to solve it efficiently, which is described as follows.
\begin{itemize}
\item In the first stage, we find a feasible solution  $\xv_{\rm f}$ to satisfy the constraints (\ref{const1})-(\ref{const3}), which is obtained by taking a solution of
\begin{equation}\label{eq:subprob1}
\onev = \mbox{sign}(\Lambdam \tilde{\xv}),
\end{equation} where $\mbox{sign}(a) = 1$ if $a\geq 0$ and $\mbox{sign}(a)=-1$ otherwise. We solve the above non-linear inverse problem efficiently via the so-called IHT algorithm. The detailed procedures are provided in Algorithm 1 where the thresholding function $[\cdot]_{\Delta}$ is defined as $[a]_{\Delta} = 1$ if $a > \Delta$, $[a]_{\Delta} = -1$ if $a \leq -\Delta$, and $[a]_{\Delta}=0$, otherwise.
\item In the second stage, we refine the above feasible solution by maximizing $\min\{\alpha_{k,1}, \alpha_{k,2}: k\in [K]\}$, which is efficiently performed using BF algorithm (see Algorithm 2). Finally, from the output of Algorithm 2 (e.g., $\tilde{\xv}$), we can obtain the proposed transmit-signal vector as $\xv = g^{-1}(\tilde{\xv})$.
\end{itemize}

{\bf Computational Complexity:} Following the complexity analysis in \cite{Li2018}, we study the computational complexity of the proposed method with respect to the number of real-valued multiplications. As benchmark methods, we consider the computational costs of exhaustive search, exhaustive search with antenna-selection, and symbol-scaling method in \cite{Li2018}, which are denoted by $\chi_{\rm E}$, $\chi_{\rm AS}$, and $\chi_{\rm S}$, respectively. From \cite{Li2018}, their complexities are obtained as
\begin{align}
\chi_{\rm E} &= 4KN_{t}\times 2^{2N_{t}}\label{eq:E-C}\\
\chi_{\rm AS}&=4KN_{\rm t}\times 3^{2N_{\rm t}}\label{eq:E-AS-C}\\
\chi_{\rm S} &= 4N_{\rm t}^2 + 24KN_{\rm t} - 2K. \label{eq:S-C}
\end{align} Recall that the proposed method is composed of the two algorithms in Algorithms 1 and 2. First, the complexity of Algorithm 1 is obtained as $t^{\star} \times 8KN_{\rm t}$ where $t^{\star} \leq t_{\rm max}$ denotes the total number of iterations. Also, the complexity of Algorithm 2, which is similar to that of refined-state in symbol-scaling method in \cite{Li2018}, is obtained as $8KN_{\rm t}$. By summing them, the overall computational complexity of the proposed method is obtained as
\begin{equation}\label{eq:P-C}
\chi_{\rm P} = 8(t^{\star} + 1)KN_{\rm t}.
\end{equation}

\begin{algorithm}
\caption{IHT Algorithm}\label{IHT}
\begin{algorithmic}[1]
\State {\em input:} $\Lambdam \in \RR^{2K \times 2 N_{\rm t}}$, $\Delta$, and $t_{\rm max}$
\State {\em initialization:} $\tilde{\xv}^{(0)} = \zerov$
\State {\em iteration:} repeat until either $\ev^{(t)} = \zerov$ or $t = t_{\rm max}$
\begin{align*}
\ev^{(t)} &= \onev - \mbox{sign}(\Lambdam [\tilde{\xv}^{(t)}]_{\Delta})\\
\tilde{\xv}^{(t+1)} &= \tilde{\xv}^{(t)} + \Lambdam^{\transp}\ev^{(t)}
\end{align*}

\State {\em output:} $\tilde{\xv}_{\rm f}  \leftarrow \tilde{\xv}^{(t)}$ 
\end{algorithmic}
\end{algorithm}



\begin{algorithm}
\caption{BF Algorithm}\label{BF}
\begin{algorithmic}[1]
\State {\em input:} $\Lambdam \in \RR^{2K \times 2 N_{\rm t}}$ and $\xv_{\rm f} \in \RR^{2N_{\rm t} \times 1}$
\State {\em initialization:} $\tilde{\xv} = \tilde{\xv}_{\rm f}$
\For{ $i=1:2N_{\rm t}$ }
\For {$j \in \{1,0,-1\}$}
\State  $\betav^{(j)} = \Lambdam \tilde{\xv}_{\{i=j\}}$ and $\beta^{(j)}_{\rm min} = \min_{i\in [2K]}\{\beta^{(j)}_i\}$
\EndFor
\State {\em update} $\tilde{x}_i \leftarrow \arg\max_{j \in \{1,0,-1\}}\{\beta_{\rm min}^{(j)}\}$
\EndFor
\State {\em output:} $\tilde{\xv} \in \RR^{2N_{\rm t} \times 1}$
\end{algorithmic}
$\Diamond$ Note that for some $j \in \{1,0,-1\}$, $\tilde{\xv}_{\{i=j\}}$ is obtained from the $\tilde{\xv}$ by simply replacing $\tilde{x}_{i}$ with $j$, i.e.,
\begin{equation*}
\tilde{\xv}_{\{i=j\} }= [\tilde{x}_{1},...,\tilde{x}_{i-1},j,\tilde{x}_{i+1},...,\tilde{x}_{2N_{\rm t}}]^{\transp}.
\end{equation*}
\end{algorithm}

\section{Simulation Results}\label{sec:SIM}

In this section, we evaluate the symbol-error-rate (SER) performances of the proposed method for the downlink MU-MIMO systems with one-bit DACs where  $N_{\rm t}=128$ and $K=16$. For comparison, we consider QZF in \cite{Saxena2017} and the state-of-the-art symbol-scaling method in \cite{Li2018} because the former is usually assumed to be the baseline approach and the latter showed an elegant performance-complexity tradeoff over the other existing methods (see \cite{Li2018} for more details). Both QPSK (or 4-PSK) and 8-PSK are considered. Regarding the proposed method, we choose the threshold parameter  $\Delta=3$ for Algorithm 1, which is optimized numerically via Monte-Carlo simulation. As mentioned in Section~\ref{subsec:model}, a flat-fading Rayleigh channel is assumed.

\begin{figure}
\centerline{\includegraphics[width=9cm]{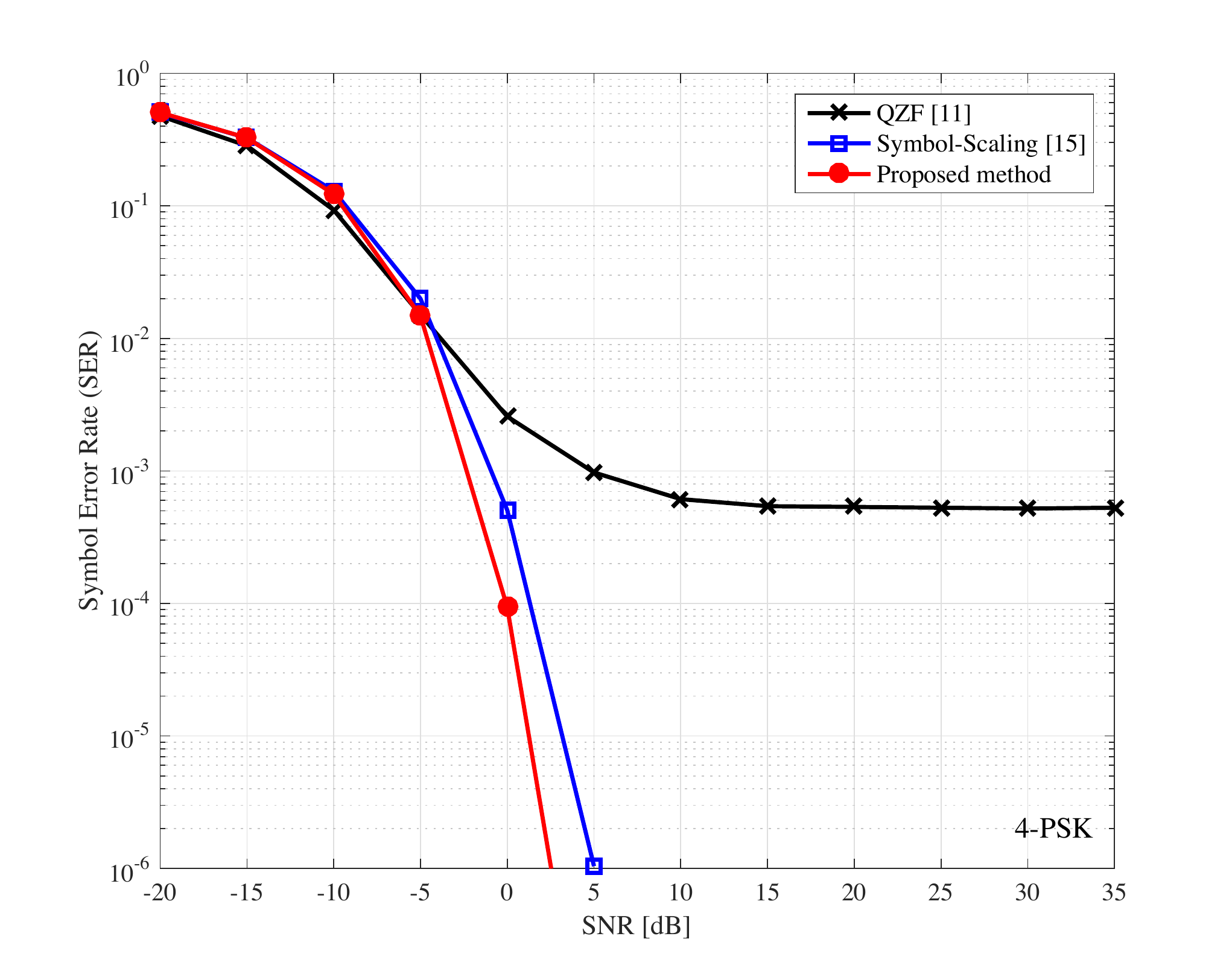}}
\caption{$N_{\rm t}=128$, $K=16$ and QPSK. Performance comparison of QZF, symbol-scaling, and the proposed methods for the downlink MU-MISO systems with one-bit DACs. The computational complexities are computed as $\chi_{\rm S}=114656$ and $\chi_{\rm P}=36044$.}
\label{sim:QPSK}
\end{figure}

Fig.~\ref{sim:QPSK} shows the performance comparison of various precoding methods when QPSK is employed. From this, we observe that QZF suffers from a serious error-floor and thus is not able to yield a satisfactory performance. In contrast, both symbol-scaling and the proposed methods overcome the error-floor problem, thereby outperforming QZF significantly. Moreover, it is observed that the proposed method can slightly improve the performance of QZF with much less computational cost (e.g., $70\%$ complexity reduction).

\begin{figure}
\centerline{\includegraphics[width=9cm]{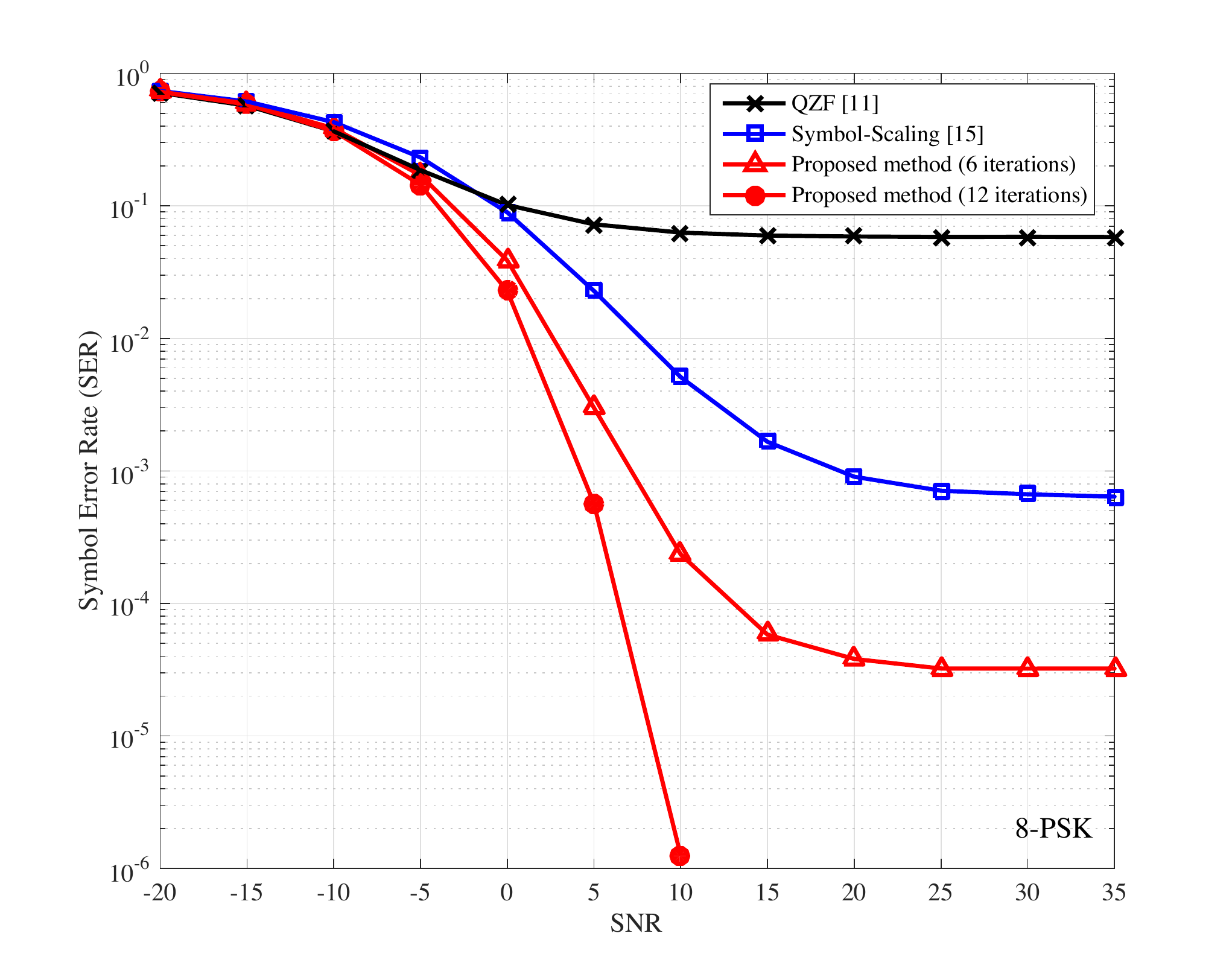}}
\caption{$N_{\rm t}=128$, $K=16$ and $8$-PSK. Performance comparison of QZF, symbol-scaling, and the proposed methods for the downlink MU-MISO systems with one-bit DACs. The computational complexities are computed as $\chi_{\rm S}=114656$, $\chi_{\rm P}=212992$ for $t^{\star}=12$, and $\chi_{\rm P}=106496$ for $t^{\star} = 6$.}
\label{sim:8PSK}
\end{figure}

Fig.~\ref{sim:8PSK}  shows the performance comparison of various precoding methods when 8-PSK is employed. It is observed that in this case, an error-floor problem is more severe than before, which is obvious since the decision regions of 8-PSK is more sophisticated than those of QPSK. Hence, antenna-selection can attain a more performance gain with a larger search-space. Accordingly, the proposed method can further improve the performance of symbol-scaling method by lowering an error-floor, which is verified in Fig.~\ref{sim:8PSK}. To be specific, the proposed method with $t^{\star}=6$  outperforms the symbol-scaling method with an almost same computational cost. Furthermore, at the expense of two times computational cost, the proposed method with $t^{\star}$ can address the error-floor problem completely. Therefore, the proposed method provides a more elegant performance-complexity tradeoff than the state-of-the-art symbol-scaling method.


\section{Conclusion}\label{sec:con}

In this paper, we showed that the use of antenna-selection can yield a non-trivial performance gain especially at error-floor regions, by enlarging the set of possible transmit-signal vectors. Since finding an optimal transmit-signal vector (encompassing precoding and antenna-selection) is too complex, we proposed a low-complexity two-stage method to directly obtain such transmit-signal vector, which is based on iterative-hard-thresholding and bit-flipping algorithms. Via simulation results, we demonstrated that the proposed method provides a more elegant performance-complexity tradeoff than the state-of-the-art symbol-scaling method. One promising extension of this work is to devise a low-complexity algorithm which is more suitable to our non-linear inverse problem for finding a feasible transmit-signal vector. Another extension is to develop the proposed idea for one-bit DAC MU-MISO systems with quadrature-amplitude-modulation (QAM). This is more challenging than this work focusing on PSK since some decision regions of QAM are surrounded by the other decision regions, which makes difficult to handle.

\end{document}